# Sprinkling Selections over Join DAGs for Efficient Query Optimization

Satyanarayana R Valluri          Soujanya Vadapalli          Kamalakar Karlapalem

**International Institute of Information Technology, Hyderabad – 500019. INDIA.**

satya@gdit.iiit.net          soujanya@gdit.iiit.net          kamal@iiit.net

## Abstract

In optimizing queries, solutions based on AND/OR DAG can generate all possible join orderings and select placements before searching for optimal query execution strategy. But as the number of joins and selection conditions increase, the space and time complexity to generate optimal query plan increases exponentially. In this paper, we use join graph for a relational database schema to either pre-compute all possible join orderings that can be executed and store it as a join DAG or, extract joins in the queries to incrementally build a history join DAG as and when the queries are executed. The select conditions in the queries are appropriately placed in the retrieved join DAG (or, history join DAG) to generate optimal query execution strategy. We experimentally evaluate our query optimization technique on TPC-D/H query sets to show their effectiveness over AND/OR DAG query optimization strategy. Finally, we illustrate how our technique can be used for efficient multiple query optimization and selection of materialized views in data warehousing environments.

## 1. Introduction

Query optimization is an age-old problem in database systems. Given a query on a set of relations it is NP-hard to find the optimal query execution strategy. One of the first and a widely popular technique is the dynamic programming formulation of access path selection by [8]. Other significant technique is the Ingres query optimizer [7]. Recently, query optimization using an integrated data structure that generates all possible ways of executing a query based on AND/OR DAG has been proven to be effective in optimizing complex queries using aggregation, group-by, and nested constructs [2]. There are some inherent properties of the AND/OR DAG that make optimal query optimization of large complex queries very inefficient:

- The generation of the AND/OR DAG for a given complex query has the time complexity $O(2^n)$, which is very large for n>10, where n is the total number of operations in a query.

- For a given number of joins, as the number of simple select conditions in the query increase, the size of the AND/OR DAG for the query increases exponentially.

- Given an AND/OR DAG, exhaustively searching for the optimal query execution plan has time complexity $O(n!)$, which is very large for n>10.

- Multiple query optimization of a set of large queries generates a very large AND/OR graph and the cost of identifying the optimal query execution plans is very prohibitive.

- In case of dynamic environments, when a new query arrives the execution plans of existing queries need to be re-constructed for multi-query optimization. This again has high complexity.

- Performing any kind of ad-hoc large query optimization requires heuristics to be applied, which can result in inefficient query execution plans.

In this paper, we present significantly new way of addressing query optimization that brings about drastic reduction in query optimization time to generate the optimal query execution plan, and can cater to multiple ad-hoc query optimization, and materialized view selection.

### 1.1 Framework

The query execution plan is modeled in terms of an AND-OR DAG. It contains all possible ways of executing a query. Given a set of queries every possible execution plan for all the queries forms an **AND-DAG**. All the possible plans together become the **AND-OR DAG**.

The union of individual **AND-OR DAG** of each query forms an AND-OR view graph. It consists of a **set of Equivalent Nodes** and a **set of Operational Nodes**. Thus the global AND-OR View Graph represents the all-possible ways of answering a given set of queries.

An **equivalent node (Eq-node)** in the DAG represents the equivalence classes of logical expression that generate the same result set, each expression being defined by a child operation node of the equivalence node and its inputs.

An **operation node (Op-node)** in the DAG corresponds to an algebraic relational operation such as 'join', `select'. It represents the expression defined by the operand and its inputs.

An eq-node can have one or more child op-nodes, each op-node being one of the possible ways of obtaining the eq-node. The existence of more than one children at any eq-node indicates an OR node. An op-node can have either one or two child eq-nodes since we consider only unary and binary operations. Every op-node represents an AND node.

A global AND-OR DAG is formed from the union of individual AND-OR DAGs of each query.

### 1.2 Related Work

The problem of query optimisation was studied extensively since the advent of database management.

Query optimisation relies on the cost model for processing queries on the database. The cost model provides the necessary formulae to estimate the intermediate results that will be generated during the execution of the query. Popular systems like System-R optimiser uses the techniques of dynamic programming and interesting orders [10], to generate near optimal query execution strategy.

A query execution plan enumerates all possible ways of executing a query. It is normally represented as a DAG. Multi-query optimisation is done by exploiting the common sub-expressions present in different queries [9]. Given a set of queries, the query plan comprehensively merges all the common operations. This plan is generated whenever a new set of queries is to be processed. [9] also describes the multi-query optimisation techniques based on materialized views. If the most commonly used node of the query plan is materialized, query processing can be made faster.

The problem of selection of materialized views is addressed in [1,3,4,5,12,13]. An algorithm based on dynamic programming searches for the most useful views to be maintained at the warehouse for the efficient query execution. In order to determine the views to be materialized all the non-leaf nodes of an AND/OR DAG are potential candidates, and hence larger the size of AND/OR DAG the more time consuming is it to select materialized views. Therefore, any method that reduces the size of AND/OR DAG without compromising on the quality of solution will make materialized view selection lot more efficient.

### 1.3 Contributions and Organization of the paper

The specific contributions of our paper are:
- A framework is presented to highlight the limitations of the **naïve solution** in optimizing queries using AND/OR graph. This framework is based on join DAGs and the concept of sprinkling selections over join DAGs to address the problem formally.
- Algorithms are presented to optimize queries using the join DAGs and results on savings in query optimization time and search space are presented for TPC-D/TPC-H queries.
- Applicability of this technique for multiple query optimization, dynamic ad-hoc query optimization, and materialized view design are presented.

The paper is organized in the following manner: Section 2 gives two motivating examples, which explain intuitively the basic idea behind the join DAG approach. Section 3 defines formally a query DAG and a join DAG. It further discusses the various algorithms for construction of complete join DAG and incremental join DAG. It ends with a short discussion on dealing with select, group by, order by, having and project operations, nested queries and complexity analysis of the problem. Section 4 gives the experimental results. Section 5 discusses the applicability of the join DAG approach for dynamic ad-hoc query processing and materialized view selection. Section 6 concludes the paper.

## 2. Motivating Examples

Consider a modified TPC-Query:

```
select n_name, sum(l_extendedprice)
from customer, orders, lineitem, supplier, nation, region
where c_custkey = o_custkey
   and l_orderkey = o_orderkey
   and l_suppkey = s_suppkey
   and c_nationkey = s_nationkey
   and s_nationkey = n_nationkey
   and n_regionkey = r_regionkey
   and r_name = 'name'
   and o_orderdate >= date '2002-02-02'
   and o_orderdate < date '2002-02-02'
group by n_name
order by l_extendedprice desc;
```

**Table1: TPC-H query (TQ1)**

The query TQ1 contains six joins and three select conditions. All possible combinations of executing these operations would be 9! = (362800) combinations. Some of these combinations might be irrelevant or redundant which would be pruned during the construction of query AND/OR DAG. For the above query given in table 1, the size of the AND/OR DAG obtained after pruning is 156 eq-nodes and 4327 op-nodes. But following our approach, the join DAG is first constructed, which contains only the join conditions. In the worst-case scenario, the join DAG might contain 6! = 720 plans since there are 6 join conditions. The select conditions are then sprinkled across the join dag at the appropriate positions to generate the optimal query execution plan. To sprinkle the selects, in this case, 720 plans need to be traversed. The size of the

AND DAG generated by join DAG approach is less compared to the naïve solution. This implies less number of plans needs to be checked to select the optimal plan. This example illustrates that the search space for the naïve solution explodes due to the combined effect of selects and joins. Isolation of selects from joins, results in immense reduction of the search space for queries with large number of joins and selects.

Separating joins and selects will prove advantageous in the case of dynamic query optimisation. When a new query arrives, in the naïve method, the global query execution plan needs to be rebuilt for multiple query optimisation. In our approach, since the join dag is pre-computed, only the new joins (if any) have to be incorporated and the optimal query execution plan searched.

The above query TQ1 took 1384 seconds to generate the best optimal execution plan using the traditional AND/OR approach on P-III Linux Server. This is due to the exponential increase in the number of possibilities at the slightest increment in the operations (join, select, project etc). For instance, if a set of queries has $j$ join conditions and $s$ select conditions. In the worst case, the number of possible plans generated would be $(s+j)!$. Hence, if $j=5$ and $s=4$, the number of possible execution plans would be $9! = (362800)$ in number. For the above single TPC query TQ1, the number of eq-nodes finally generated after pruning irrelevant combinations, is 156 and the number of op-nodes generated is 4327.

While in the case of Incremental Join DAG, it takes 24 seconds to compute the best execution plan. The method applied is to first generate the global multiple execution plan only for the various joins occurring in the queries instead of constructing for all the joins possible between the various relations.

When a new query TQ2 was considered, the number of eq-nodes generated increased drastically, just for an addition of one select operation.

```
select l_returnflag, l_linestatus, sum(l_quantity),
       sum(l_extendedprice), avg(l_quantity),
       avg(l_extendedprice), avg(l_discount), count(*)
from lineitem
where l_shipdate <= date '1998-12-01'
group by l_returnflag, l_linestatus
order by l_returnflag, l_linestatus;
```

**Table2: TPC-H query TQ2**

The number of eq-nodes generated after reconstructed the whole AND/OR DAG in the naïve method is 277 and the number of op-nodes is 4448. To incorporate one more select condition into the AND/OR DAG the naïve method generates 277-156=121 more eq-nodes, while in the join DAG approach, only 6 eq-nodes are created.

The above discussion illustrates that the join DAG approach proves to be more efficient than naïve method in terms of space and time complexity (discussed in section 3.6 and 3.7).

## 3. Query DAGs and Join DAGs

The join conditions in a database schema can be represented in the form of a database schema graph [6]. It is a connected graph, where each node is a relation and a hyper edge between two nodes $Ri$ and $Rj$ gives the possibilities of joins between $Ri$ and $Rj$.

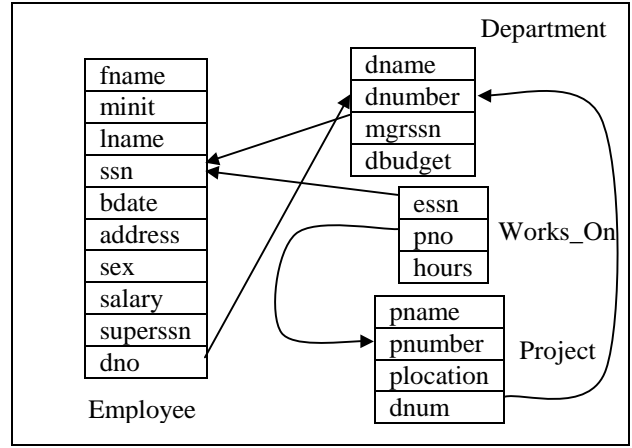

**Figure 1: Example database**

Figure 1 shows one of such database schema graph. Using the database schema graphs the foreign key – primary key relationships between various relations can be extracted to construct a query DAG.

Formally, a query DAG can be defined as: QD = <$E$, $O$, $A_e$, $A_o$, $R_e$> where $E$ is the set of eq-nodes, $O$ is the set of op-nodes, $A_e$ is a set of directed arcs between eq-nodes ($E$) and op-nodes ($O$), $A_o$ is a set of directed arcs between op-nodes ($O$) and eq-nodes ($E$) and $R_e$ is the set of eq-nodes such that:

- every $o \in O$ where o represents a unary operation like select contains one child eq-nodes, child($o$) ($o$, child($o$)) $\in A_o$.
- every $o \in O$ where $o$ represents a binary operation like join contains two child eq-nodes left child, child$_l$($o$) and right child, child$_r$($o$), ($o$, child$_l$($o$)) $\in A_o$ and ($o$, child$_r$($o$)) $\in A_o$.
- $k$ children of $e \in E$ child$_i$($e$) where $i = 1$ to $k$ represent the $k$ possible ways of obtaining $e$ and ($e$,child$_i$($e$)) $\in A_e$.
- for every $o \in O$ and ($o$, child($o$)) $\in A_o$ and $e \in E$ such that ($e$, child($e$)=$o$) $\in A_e$, defn($e$), definition of $e$ represents the result of unary operation $o$ performed on child($o$) that is defn($e$)=$o$(child($o$)).
- for every $o \in O$ and ($o$, child$_l$($o$)) $\in A_o$ and ($o$, child$_r$($o$)) $\in A_o$ and $e \in E$ such that ($e$, child($e$)=$o$) $\in A_e$, defn($e$), definition of $e$ represents the result of binary operation $o$ performed on child$_l$($o$) and child$_r$($o$) that is defn($e$) = $o$(child$_l$($o$),child$_r$($o$)).

- size of $e \in E$ represents the size of the node obtained by applying $o \in O$ over $e$ where $(e,o) \in A_e$.
- cost of $o \in O$ is the cost of operation applied by $o$.
- $R_e = <e_1, e_2, \ldots e_q>$ where $e_i$ represents the eq-node corresponding to query $q_i$.

A join DAG is a special case of query DAG where every $o \in O$ represents a join operation and contains exactly two children $child_l(o)$ and $child_r(o)$ such that $(o, child_l(o)) \in A_o$ and $(o, child_r(o)) \in A_o$. A join DAG is built based on the join conditions of the queries.

Join DAG is formed using the join combinations, which represent the various possible ways of executing the joins. For the above set of joins, in query TQ1, **J1** to **J6**, the total number of join combinations possible is 6! =720. But some of the join combinations may be identical, for example, for a chain query: A $J_{A,B}$ B $J_{B,C}$ C $J_{C,D}$ D, the join combinations $J_{A,B}$, $J_{B,C}$, $J_{C,D}$ would be same as $J_{C,D}$, $J_{B,C}$, $J_{A,B}$ since $J_{A,B}$, $J_{C,D}$ can be performed parallel. The reduction in join combinations for AND/OR DAG and join DAG are the same, therefore, performing worst-case comparison shows the same trends as the experimental results.

There are two types of join DAGs: Complete History Join DAG and Incremental History Join DAG. For query optimisation it does not matter whether we use complete history join DAG or incremental join DAG, once they are constructed. From section 3.3 onwards, we assume that the join DAG is available and use the term join DAG to mean either complete history join DAG or incremental join DAG.

### 3.1 Complete History Join DAG

A complete history join dag represents all possible ways of applying the known join conditions between various relations. This join dag is computed once and can be stored for future access, as the possible ways of performing a join-order does not change.

To generate the complete history join DAG, all the foreign key relationship constraints of the schema are noted and various join combinations are generated. After applying the each join combination, the common plans are merged. Pseudo code for the generation of complete history join DAG is given in table 3.

If the complete history join DAG is represented as an AND/OR DAG, then the root corresponds to the eq-node corresponding to the conjunction of all the join conditions in J.

### 3.2 Incremental Join DAG

Though complete history join DAG stores the join conditions is a concise manner, if the incoming queries do not use all the join conditions, it contains useless plans, which will not be explored for any queries. The solution for this is to store the join combinations of the join conditions, which are being used by the present set of queries. The algorithm for the generation of incremental join DAG is shown in table 4.

```
Input: Set of join conditions J=<j₁, j₂…jₙ> and the
set of base tables T
Output: Complete history join dag CHJ=<Eₕ, Oₕ,
Aₑₕ, Aₒₕ, Rₕ>
Procedure:
begin
  set Eₕ, Oₕ, Aₑₕ, Aₒₕ to empty
  form join combinations C=<C₁, C₂,…Cₘ> from
  J
  for each join combination Cᵢ=<jᵢ₁, jᵢ₂,…jᵢₙ> in C
    for each jᵢₖ∈ Cᵢ do
      if ∃ o∈ Oₕ such that o=jᵢₖ then
        set oₙ=o
      else
        create a set of op-nodes Oₘ such
        oₙ∈ Oₘ such that o=jᵢₖ
        set childₗ(o) and childᵣ(o) to the joining
        nodes of jᵢₖ.
      end if
      if ∃ eₖ∈ Eₕ, (eₖ, child(eₖ)=o) ∈Aₑ and o∈ Oₕ
      such that o=jᵢₖ then
        if oₙ! = o then
          add o as child of eₖ that is (eₖ, o)
          ∈ Aₑ.
        else
          create eₙ∈ Eₕ and make (eₙ, oₙ) ∈ Aₑ
          such that child(eₙ)=oₙ
        end if
      end if
    end for
end
```
**Table 3: Algorithm for generation of complete history join dag**

```
Input: Old history join dag OldHJ=<Eₒ, Oₒ, Aₑₒ,
Aₒₒ,Rₒ> and a set of join conditions J=<j₁, j₂…jₙ>
Output: New history join dag NewHJ=<Eₙ, Oₙ,
Aₑₙ, Aₒₙ,Rₙ> such that it contains J.
Procedure:
begin
  copy OldHJ into NewHJ
  for each join condition jᵢ in J do
    if ∃ o∈ Oₙ such that o=jᵢ then
      set oₙ=o
    else
      create a set of op-nodes Oₘ where ∀ o∈ Oₘ
      such that o=jᵢₖ
      set childₗ(o) and childᵣ(o) to the joining
      nodes of jᵢₖ.
    end if
    if ∃ eₖ∈ Eₙ, (eₖ, child(eₖ)=o) ∈Aₑₙ and o∈ Oₙ
    such that o=jᵢₖ then
      if oₙ! = o then
        add o as child of eₖ that is (eₖ,o) ∈Aₑₙ.
      end if
```

```
        else
            create eₙ∈Eₙ and make (eₙ, oₙ) ∈ Aₑₙ such
            that child(eₙ)=oₙ
        end if
    end for
end
```

**Table 4: Algorithm for generation of incremental history join dag**

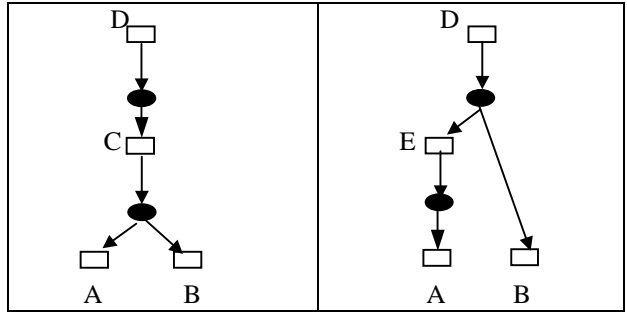

**Figure2: Order between joins and selects**

### 3.3 Generation of Global Query Execution plan

To generate the query execution plan from the join DAG thus obtained from either of the above two methods, the other operations are inserted in the join DAG, to make the plan complete with respect to the operations required in the queries.

#### 3.3.1 Sprinkling Select operations

Each select operation is performed on every eq-node starting from the base eq-node to the root node in the join DAG.  For every possible placement of the select operation, the cost of execution of that corresponding plan is checked. If the cost of execution is minimum, then the select is inserted at that placement. We call this technique of identifying the optimal placement of select conditions as *sprinkling select operations on join AND/OR DAG*. Note that while sprinkling select operations the costly plans (as determined by selection selectivity factors) are pruned to make sprinkling process efficient.

To decide the order between a select and a join, we use a simple empirical formula. The figure shows the two possible ways of applying a join condition and a select condition. The cost of first method is cost1 $|A|*|B|+|C|$ where $|node|$ = size of node. The size of C is estimated using the join selectivity factor using the formula:

$$|C| = jsf * |A| * |B|$$

The cost with second method is cost2 $|A|+|E|*|B|$ where $|E| = ssf *|A|$.

If cost1 < cost2 (case 1), then applying select operation after join operation would be beneficial. On the other hand, if cost1 > cost2 (case 2), then pushing down the select operation and applying join after select would be beneficial. Depending on which condition is applicable, we decide the order between the join and the select operation. The complete algorithm for applying selects is given in table 5. Note that statistics collected in the system catalog are used to identify the optimal sprinkling of selects in a join DAG.

When a new query contains a new join condition, it is incrementally added to the join DAG. Thus, the incremental join DAG is computed as and when new join orders are detected. These new operations are incorporated into the history join-dag, thus making it up-to-date with the current possible join-orders.

```
Input: Join DAG JD=<Eⱼ, Oⱼ, Aₑⱼ, Aₒⱼ, Rⱼ> and a set
of select conditions, S=<s₁, s₂,….sᵣ>
Output: New DAG ND=<Eₙ, Oₙ, Aₑₙ, Aₒₙ, Rₙ> with
joins and select conditions
Procedure:
begin
    copy JD to ND
    for each select condition sᵢ in SC do
        let bᵢ∈Eₙ be the base table on which sᵢ is to
        be applied and set tᵢ=bᵢ
        let oⱼ∈Oₙ be such that (oⱼ, tᵢ) ∈Aₒₙ and (oⱼ,
        bₖ) ∈Aₒₙ and (eₗ, oⱼ) ∈Aₑₙ --- step(i)
        val1 = |tᵢ|*|bₖ|+eₗ
        val2 = |tᵢ|+(ssf)*|tᵢ|*|bₖ| where ssf represents
        the selectivity factor of sᵢ retrieved from
        system catalog
        if (val1>val2) then
            create oₛ∈Oₙ and add (oₛ, tᵢ) to Aₒₙ and
            create eₜ∈Eₙ and add (eₜ, oₛ) ∈Aₑₙ
            remove (oⱼ, tᵢ) and add(oⱼ, eₜ) to Aₑₙ and
            update ND.
        else
            set tᵢ=eₗ
            if ti is not root node then
                repeat step (i)
            else
                create oₛ∈Oₙ and add (oₛ, eₗ) to Aₒₙ
                and create eₜ∈Eₙ and add (eₜ, oₛ)
                ∈Aₑₙ
            end if
        end if
    end for
end
```

**Table 5: Algorithm to Sprinkle Selects**

#### 3.3.2 Sprinkling Project operations

The above method for select operations is applied for the project operations. The project operations have to be placed over the appropriate node in the join DAG. The attributes which are required by the higher level parents are projected at every stage. But these may interfere with the query optimisation since same plan can be shared by more than query. Depending on whether it is single query optimisation or multiple query optimisation the attributes needed for all queries being optimised are appropriately projected.

### 3.3.3 Sprinkling Group by and having operations

The same algorithm given in table 3 can also be extended for group by operation also. The formulae to calculate *val1* and *val2* are:

$val1 = |t_i|*|b_k|+e_1$
$val2 = |t_i|+d*|b_k|$

where $d$ = number of distinct values which the group by attribute can take.

The above formula calculates whether it is beneficial to apply a group by before an operation or after an operation. The having operation is similar to the select condition except that the having operation is to be performed after the group operation is applied and thus the search space still becomes smaller.

### 3.3.4 Sprinkling Order by operations

The formulae for order by operation would be:

$val1 = |t_i|*|b_k|+e_1$
$val2 = |t_i|+|t_i|*|b_k|$

The result of the operation after applying an order by condition will be same as the table itself since the tuples are ordered based on the order attribute.

### 3.4 Nested Join Queries

Nested queries are optimised by recursively calling the join DAG algorithm and merging results. That is, for a two level query, the inner query is first optimised using the above algorithm. After that the outer query is optimised. These two query plans are merged to generate optimal query plan.

### 3.5 Examples

Consider the Company schema and the foreign key references shown in figure 1.

**Query1**: Give the first name and last name of the employees and the name of the project they work on for more than 30 hours and which is located at 'Hyderabad'

**SQL Query**:
Select fname,lname,pname
From Employee, Project,Works_On
Where Employee.ssn = Works_On.essn
And  Works_On.pno = Project.pnumber
And  Works_On.hours > 30
And  Project.plocation = 'Hyderabad'

**Query2**: Give the first name and last name of the employees, the name of the project in which they work for more than 30 hours which are located in 'Hyderabad' and the name of controlling department whose budget is more than 30k.

**SQL Query**:
select fname,lname,dnumber,hours,pname
from Employee , Project, Department, Works_On
where Employee.ssn = Works_On.essn
and  Works_On.pno = Project.pnumber
and  Project.dnum = Department.dnumber
and  Works_On >30
and  Project.plocation = 'Hyderabad'
and  Department.dbudget > 30k

**Table 6: Example queries**

Table 6 shows two example queries on the database described in table 5. Figure 3 shows the complete join DAG for query 1.

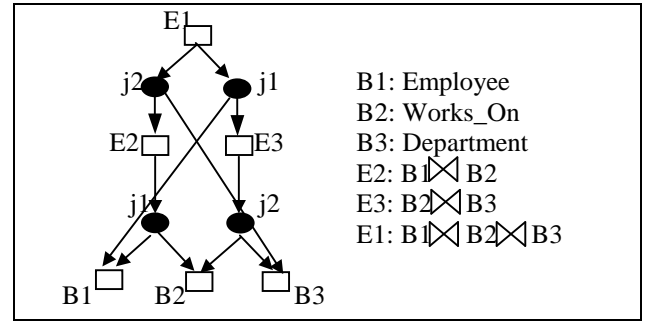

**Figure 3: Complete Join DAG for query 1**

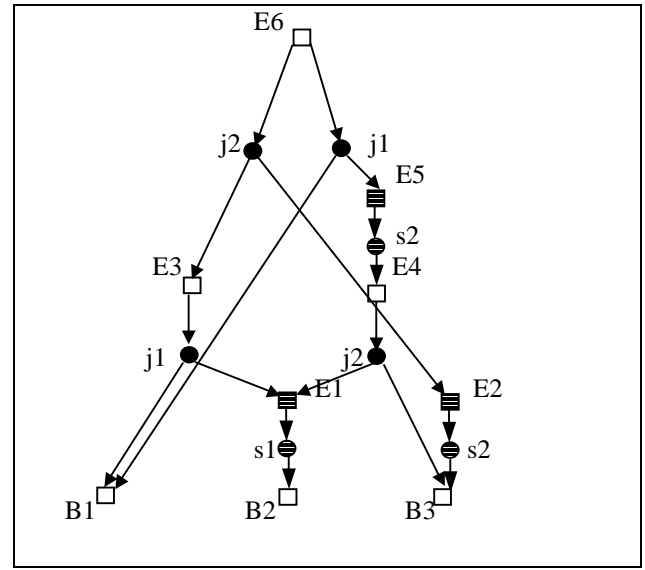

**Figure 4: Final DAG for query1**

Query 1 contains a select condition hours>30 which can be applied on three nodes in the tree, B2, E2, E3 and E1. We use the formulae developed in section 5.3.1. to decide the exact location where the select condition is to be applied. Figure 4 shows the AND/OR DAG generated by sprinkling selects over the join DAG. For the same query in the exhaustive method, we need to consider 4! = 24 join combinations, where as using our approach, we need to consider only 2! = 2 combinations. In the join DAG approach, even after the insertion of the select condition, the search space will not increase since the same the new select condition is inserted in the existing two plans. For the query 2, since we have 3 join conditions, we need to consider 3! = 6 combinations, where as in the naïve solution we need to consider, 6! = 720 combinations. In the complete join DAG method, the history join DAG is built based on the foreign key relationships known between the relations, which are 5 for the example schema. So the join DAG has to consider 120 plans. Though this cost of considering 120 plans may be amortized over a period of time when a set of queries are processed, this may not be beneficial for the query processing of only these two queries. The solution for this problem would be to construct the history join DAG incrementally for the new join conditions given by the new queries. Thus, we construct the history join DAG for queries shown in table 6. Then when query 2 arrives, we see the new join condition and add the new nodes to the history join DAG. Thus, history join DAG method can be used for dynamic query optimisation also. In case of single query optimisation, since the join conditions of the queries are already dealt with offline and stored, only the select conditions need to handled. In case of multiple queries, the common sub-expressions that are existent between them are captured in the join DAG. Thus the join DAG approach can be used for both single query and multiple query optimisation.

### 3.6 Space Complexity

This section deals with the estimation of number of eq-nodes generated in join DAG and AND/OR DAG with an insertion of a new select condition.

Consider a join DAG, with level $n$ (i.e. $n$ joins) and $p$ number of plans. A select operation is checked with all the nodes in a plan (i.e $n+1$ nodes) and is placed over that node, which leads to minimum query execution cost. Hence, upon each plan a search is performed to choose that node over which select operation could be done in order to reduce the cost. After the insertion of a select operation, the level becomes $n+1$ because it would now mean ($n$ joins + 1 select) operations. Hence after insertion of s selects, the number of nodes in a plan is ($n + s$) nodes.

Hence, for insertion of $q$ select operations into a single plan of a join DAG would be:
In the worst case, all the select conditions are applied over the same relation.
After the insertion of 1$^{st}$ select:
  No of plans = $p$
  Level of each plan = $n + 1$
  No of eq-nodes = increase by $p$ = $N_{eq} + p$
  where $N_{eq}$ is the number of eq-nodes
After insertion of 2$^{nd}$ select:
  No of plans = $p$
  Level of each plan = $n + 2$
  No of eq-nodes = increase by $p$
      = $(N_{eq} + p) + p$
      = $N_{eq} + 2p$
After inserting of qth select:
  No of plans = $p$
  Level of each plan = $n + q$
  No of eq-nodes = increase by $p$

      = $(N_{eq} + q*p)$

### 3.6.1 Estimating number of eq-nodes and plans

We need to estimate the number of eq-nodes and number of plans in And/Or DAG with an insertion of a new select condition.

Consider an and/or dag, with level $n$ (i.e. a total of $n$ operations (join operations + select operations). When a new select operation needs to be inserted, it would imply that each plan in turn would lead to $n$ different plans. This is because the select operation could be performed on any of the nodes that are parents of the base table upon which the select needs to be done.

Hence, after the insertion of one select operation:
  No of plans = $p * n$
  Level of each plan = $n + 1$;
  No of eq-nodes = $N_{eq} + n * p$
After insertion of 2$^{nd}$ select operation:
  No of plans = $(p * n) * (n + 1)$
  Level of each plan = $n + 2$;
  No of eq-nodes = $N_{eq} + n * p + (n + 1)*(p*n)$
After insertion of qth select operation:
  No of plans = $p * n * (n+1) * (n+2) * \ldots *(n+q-1)$

$$= p * \prod_{i=0}^{s-1}(n+i)$$

Level of each plan = $n + q$
No of eq-nodes
= $N_{eq} + n*p + (n+1)*n*p + \ldots + (n+q-1)(n+q-2)\ldots(n+1)(n)*p$

$$= Neq + p * \sum_{k=0}^{q-1}\prod_{i=0}^{k}(n+i)$$

### 3.7 Time Complexity

The complexity of traditional AND-OR DAG generation is given below.
Let Q be the query, which contains
  s = number of select conditions
  j = number of join conditions
  Total number of operations: n = s + j

| Merging Plans | No. of operations | No. of resulting plan |
|---|---|---|
| 1, 2 | n * n | n + n/2 = 3n/2 |
| 3 | 3n/2 * n | 3n/2 + n/2 = 2n |
| 4 | 2n*n | 2n + n/2 = 5n/2 |
| 5 | 5n/2*n | 5n/2 + n/2 = 3n |
| …. | …. | …. |
| n! | n!*n/2 * n | n!*n/2 + n/2 = (n!+1)*n/2 |

**Table 7: Complexity of merging n! plans**

To build an AND-OR DAG, we need the following steps:
*Step 1*:

We need to generate all possible permutations of the conditions, which will be (s+j)! = n!

*Step 2*:

Merging all the valid plans into a single AND-OR DAG. In the worst case, all the plans are valid plans, and hence need to merge all the plans. Each plan contains n number of operations (op-nodes). The complexity is summarized in the table 7.

We assume that on an average the number of nodes of the resulting plan formed by merging plan1 and plan2 is no_of_nodes(plan1)+(no_of_nodes(plan2)/2). That is, we assume that there is 50% overlap between the two plans.

Therefore the total number of computations involved in generation and merging of eq-nodes and op-nodes while constructing the and/or DAG is:

= $n*n + 3n/2*n + 4n/2*n + 5n/2*n + \ldots\ldots +n!*n/2*n$

= $n^2/2 [ 2+3+4+5+\ldots\ldots+n!]$

= $n^2/2 [ (1+2+3+4+5+\ldots\ldots+n!)-1]$

= $n^2/2 [ (n! * (n!+1)/2) -1]$

So, the total complexity is:

= $n! + n^2/2 [ (n! * (n!+1)/2) -1]$

**Complexity of Best Execution Plan Generation (from Join DAG and sprinkling selects)**

Following the same notation, the number of join conditions is j and the number of select conditions is s.

To generate all the possible join DAGs from the join conditions, in the worst case the complexity would be j!

In the above-generated join DAG, to merge the common nodes among the various plans the total complexity would be (as done for AND-OR DAG above)

$j^2/2 [ (j! * (j!+1)/2) -1]$.

To order the select conditions according to the measure - frequency per tuple, the complexity would be of $s^2$.

To insert the selects into the join DAG, the complexity would be j(j!+1)/2 * s.

A select can be placed anywhere in a plan from the root node to the base table on which the select is to be operated. Hence, if there are, say *m* join operations in a plan, then the current select condition need to be compared with all the join selectivity factors i.e. the possible size of the node and check whether inserting the select condition over the join is beneficial. Hence, every select condition needs to be compared j times and the select is placed in that position, which ensures minimum cost of query execution.

In the worst case, all the selects need to be placed just above the base table node, i.e. selects have to be pushed down. This would require comparison of all selects with all the joins in the plan. The complexity thus, would be (j(j!+1)/2)*s

For all the possible join plans in the join DAG, each select has to be compared with j join conditions. Hence for all selects it is j*s. This has to be done for all the plans possible, which turns out to be j * s * j!

Hence the total complexity comes up to

$j! + j^2/2 [ (j! * (j!+1)/2) -1] + (s^2)+ (j * s * j!)$

For example, if we take j = 4 and s = 3

Comparison between naïve approach and join DAG approach: Consider a query with 4 joins and 3 select operations. In the worst case, complexity to build a traditional and-or dag would be 6356724. While in the case of the join DAG approach, it would be 2713. Hence, generation of the global multiple execution plan would be much faster in case of join DAG approach in comparison with the naïve approach.

## 4. Experimental Results

The experiments were performed over two datasets, one over a company schema and another on a sub-set of modified TPC queries.

### 4.1 Implementation Details

The algorithms presented in Section 3 are implemented in C language on linux platform. The naïve solution and the join DAG method programs were about 10k lines each. The results are evaluated using a company database schema a part of which is presented in figure 1. A subset of TPC-D/H queries is also evaluated.

### 4.2 Empirical Results

The set of TPC-Queries over which the tests are performed, are shown in table 8.

| **Q1**: select l_returnflag, l_linestatus, sum(l_quantity), sum(l_extendedprice), avg(l_quantity), avg(l_extendedprice), avg(l_discount), count(*) from lineitem where l_shipdate <= date '1998-12-01' group by l_returnflag, l_linestatus order by l_returnflag, l_linestatus; |
|---|
| **Q2**: select l_orderkey, sum(l_extendedprice), o_orderdate, o_shippriority from customer, orders, lineitem where c_mktsegment = 'mkt1'     and c_custkey = o_custkey     and l_orderkey = o_orderkey     and o_orderdate < date '2002-02-02'     and l_shipdate > date '2002-02-02' group by l_orderkey, o_orderdate, o_shippriority order by l_extendedprice desc, o_orderdate; |
| **Q3**: select n_name, sum(l_extendedprice) from customer, orders, lineitem, supplier, nation, region |

```
where c_custkey = o_custkey
    and l_orderkey = o_orderkey
    and l_suppkey = s_suppkey
    and c_nationkey = s_nationkey
    and s_nationkey = n_nationkey
    and n_regionkey = r_regionkey
    and r_name = 'name'
    and o_orderdate >= date '2002-02-02'
    and o_orderdate < date '2002-02-02'
group by n_name
order by l_extendedprice desc;
Q4: select sum(l_extendedprice)
from lineitem
where l_shipdate >= date '2002-02-02'
    and l_shipdate < date '2002-02-02'
    and l_discount < 20 and l_discount > 0
    and l_quantity > 3;
```

**Table 8: TPC-H queries used for evaluation**

Table 9 shows the sizes of the and-or DAGs generated from both the approaches.

| | MODIFIED TPC/H QUERIES | | | |
|---|---|---|---|---|
| | *ANDOR* | | *join DAG* | |
| | Eq-Nodes | Op-Nodes | Eq-Nodes | Op-Nodes |
| Q1 | 7 | 6 | 7 | 6 |
| Q2 | 28 | 53 | 16 | 14 |
| Q3 | 33 | 81 | 7 | 6 |
| Q4 | 154 | 679 | 88 | 156 |

**Table 9: Comparison of naïve approach vs join DAG approach for TPC/H queries**

Figure5 and Figure6 represent the performance evaluation of queries for another schema 'company schema'. Figure5 depicts the ratio of the number of eq-nodes generated from the naïve approach to that of our approach, with the inclusion of a new query every time to the existing set of queries.

While the X-axis shows the number of join conditions, Y-axis shows the ratio where ratio is the number of equivalent nodes generated from naïve approach to the number of equivalent nodes generated from the join DAG approach. It shows the rise in the number of eq-nodes being generated with every new join operation being encountered in the query. The join DAG approach is more efficient as the number of join operations increase. While the X-axis shows the number of join conditions, Y-axis shows the ratio where ratio is the number of equivalent nodes generated from naïve approach to the number of equivalent nodes generated from the join DAG approach. It shows the rise in the number of eq-nodes being generated with every new join operation being encountered in the query. The join DAG approach is more efficient as the number of join operations increase.

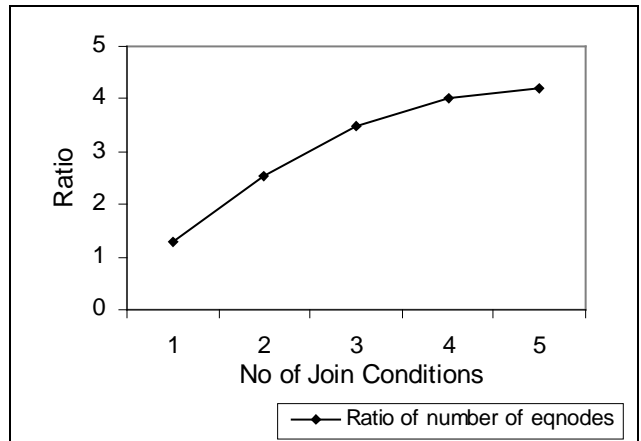

**Figure 5: Ratio of Eq-Nodes vs No Of Join Operations**

In the same manner, a graph is plotted for the number of op-nodes being generated from both the approaches. This is shown in Figure 6. Here, X-axis shows the number of join conditions and the Y-axis shows the ratio of number of op-nodes generated by naïve method to the number of op-nodes generated by the join DAG approach.

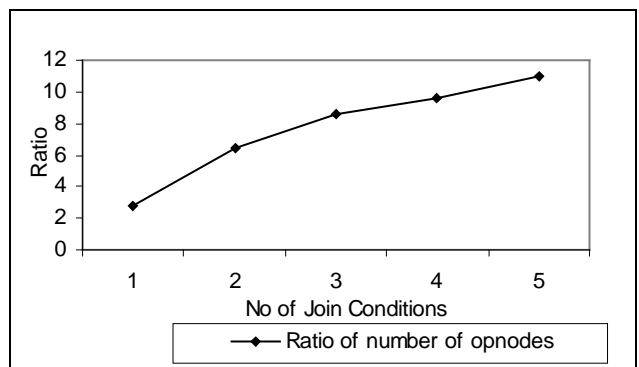

**Figure 6: Ratio of Op-Nodes vs No Of Join Operations**

For every new join condition, the naïve method generated many new plans, which result in generation of too many op-nodes which is avoided in the join DAG approach since each new operation is applied at the appropriate place.

Figures 5 and 6 check the performance of both approaches when a new join is encountered. The same has been done for the varying number of select operations, i.e. how do the size of the final DAG generated (number of eq-nodes and op-nodes) vary with a new select condition encountered. Figure 7 and 8 depict the same.

From figures 7 and 8, one can observe that as the number of select conditions increase, the slope of the graphs becomes more steep which shows that in the naïve solution the number of plans increase exponentially with the number of select conditions. The same inference can also be verified by the estimation curves shown in figure 10 and 12.

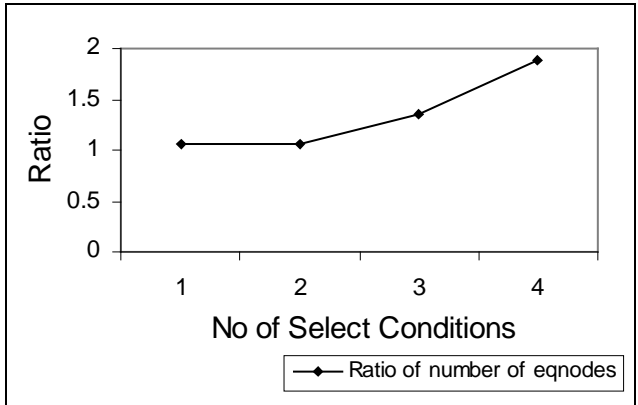

**Figure 7: Ratio of Eq-nodes vs No Of Select Operations**

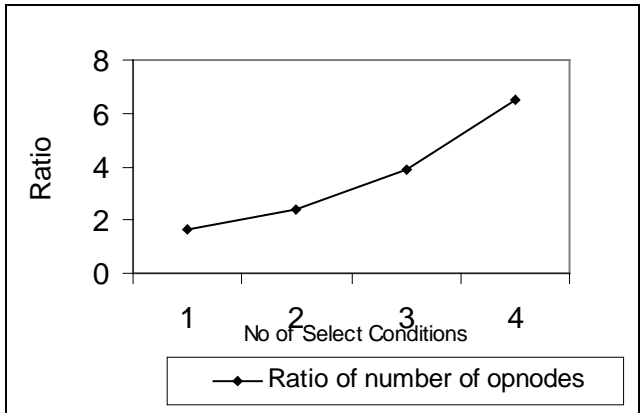

**Figure 8: Ratio of Op-nodes vs No Of Select Operations**

### 4.3 Analytical Experiment Results

Due to lack of sufficient memory on the machine, we have done an estimation of number of plans using the time and space complexity formulae given in section 3.6 and 3.7. The analytical experimental results consist of:
- Comparison of eq-nodes generated for a single query/multiple queries for naïve approach vs join DAG approach for varying number of joins and selects.
- Comparison of time complexity in generating the optimal plan for naïve approach vs join DAG approach.

Figures 9 and 10, are the graphs that depict the relative performance evaluation of both the approaches in the case of a single query, when the number of joins and selects operations vary.

From figures 9 and 10, one can infer that the increase in the slope of the curve as the number of select conditions increase is higher when compared to the increase in the slope of the curve as the number of join conditions increase.

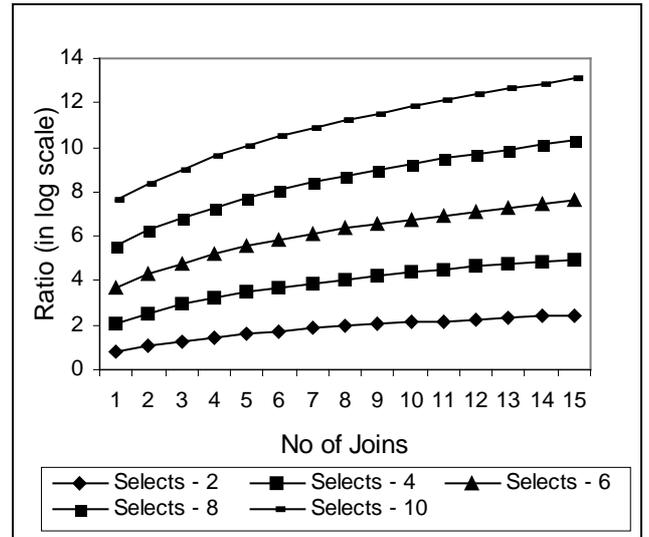

**Figure 9: Ratio of number of eq-nodes of DAG for single query with varying joins**

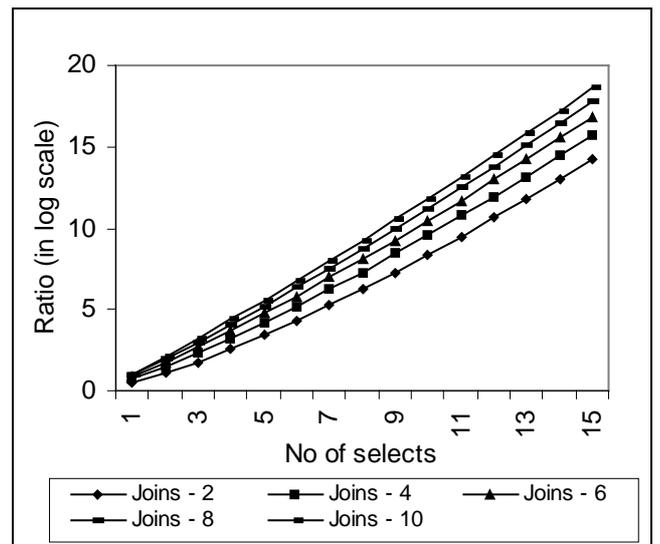

**Figure 10: Ratio of number of eq-nodes of DAG for single query with varying selects**

This is because, with increase in the number of join conditions, along with the size of AND/OR DAG in the naïve method, the size of join DAG also increases to incorporate the new join condition. But, with the increase in the number of select conditions, the size of join DAG remains constant while the size of AND/OR DAG increase exponentially.

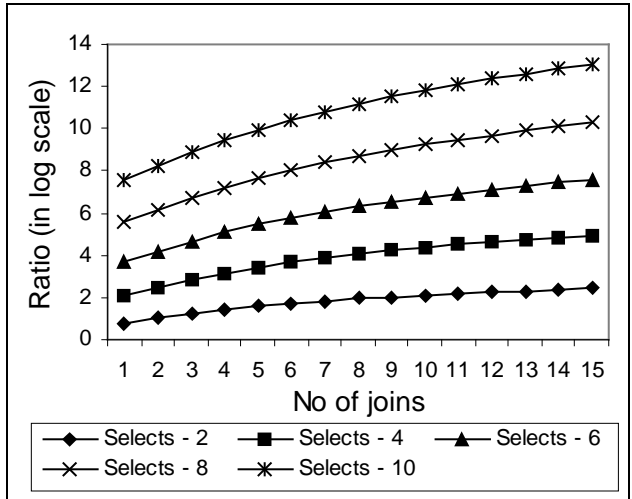

**Figure 11: Ratio of number of eq-nodes of DAG for MQO with varying joins**

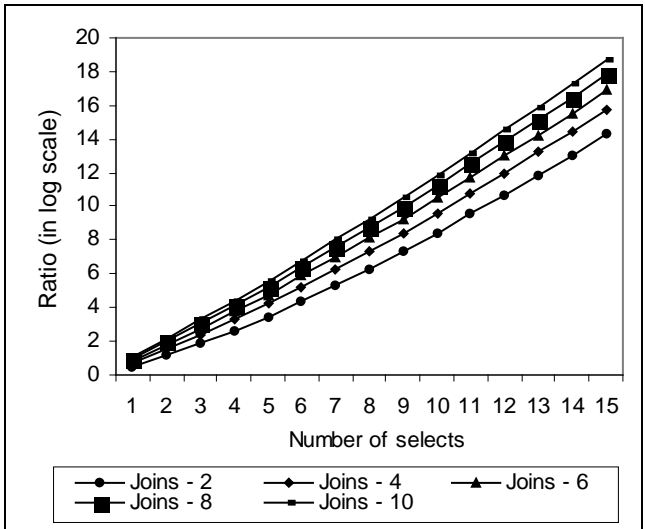

**Figure 12: Ratio of number of eq-nodes of DAG for MQO with varying selects**

Figures 11 and 12 show the estimation curves for the ration of number of eq-nodes of DAG for multiple query optimisation with varying joins and selects. The steep in the curve for figure 12 can be observed.

The figure 13 shows that as the number of joins and the number of selects in a query increase, the join DAG approach, is substantially more efficient than the naïve solution.

## 5. Applications

### 5.1 Materialized View Selection

The problem of materialized view selection can be defined as selecting a subset of intermediate nodes of the query execution plan which help in speeding up the process of query answering, subject to constraint on resources like storage space, cost of view maintenance [3,4,5].

Materialized views can be selected on the AND/OR DAG generated by join DAG approach. Since the AND/OR DAG generated by join DAG approach will be smaller than that of naïve solution, the search space for materialized view selection gets reduced significantly, without compromising on the optimal solution. This is similar to the MVPP heuristic proposed in [12,13].

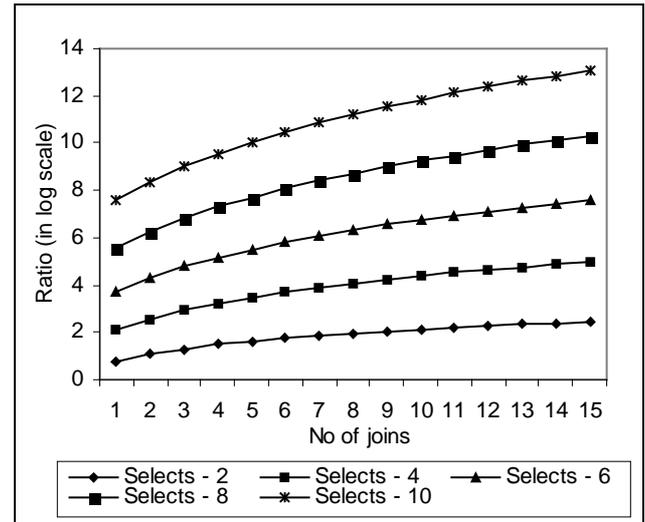

**Figure 13: Ratio of Time Complexity of Optimal Plan Search**

### 5.2 Dynamic Ad-Hoc Queries

Most of the practical applications like OLAP involve ad-hoc complex queries, which are dynamic in nature. The solutions proposed for query optimisations normally deal with a fixed set of queries where the statistics about the queries like frequency of execution are available. If these are extended for dynamic queries, they incur a lot of cost since the query execution plans for the queries need to be recomputed. The join DAG approach solves this problem since the query execution plan for join conditions is already computed.

**Single query optimisation**: When a new ad-hoc query is to be executed, the appropriate eq-node for the join conditions of the query is selected, and for all plans from this eq-node to the leaves nodes appropriate, select conditions are placed. The placement that gives minimal cost gives the optimal query execution plan.

**Multi-query optimisation**: In this case, a set of queries is already being executed when the new query starts to execute. Therefore, it is possible that some of the join conditions (or all of the join conditions) in the query are already present in the multiple queries AND/OR DAG. If so, the best plan among them is selected. Otherwise the join conditions of the new query are augmented to the history DAG and then selects of the new query are sprinkled to identify the best plan to be used for all the queries being executed. Since the history DAG is partially pre-computed, only the cost of identifying the best plan has to be borne. This reduces the time for ad-hoc multiple query optimisation substantially.

**With materialized view support**: Materialized view selection depends on the creation of multiple query AND/OR DAG, and searching for the set of nodes to be materialized for efficient query execution. This requires the complete AND/OR DAG. But by using the join DAG we get a compact structure of different join orders, while sprinkling the selects determining the effectiveness of materializing that view is performed. This technique enables a cost-effective way of generating multiple query AND/OR DAG, and efficient search procedure for identifying the views to be materialized for ad-hoc queries.

## 6. Conclusion and Future work

Query optimisation is an age-old problem in database systems. In this paper, we have enhanced the AND/OR DAG based query optimisation by (i) pre-computing the join AND/OR DAG for identifying the eq-node for the join-conditions in a query, and (ii) sprinkling the selections along the each query plan to identify the optimal query execution plan. The two main advantages of this technique are (i) the join DAG is computed once and is made persistent with an index to efficiently retrieve parts of it for query optimisation, and (ii) instead of searching many multiple plans for determining the optimal query execution plan in naïve solution, the sprinkling of selections drastically reduces the number of query execution plans to be searched and processed. This significant technique has been implemented and tested against the naïve AND/OR DAG query optimisation. The results show a significant reduction in size of the AND/OR DAGs, the time taken to generate the best query execution plan for a query when the number of join and select conditions in the query increase. The utility of this technique for optimizing single or multiple ad-hoc queries and selecting materialized views is presented.

Future work involves applying this technique for optimising large costly queries in data warehouing environments, and integrating the optimiser within a public domain relational DBMS.